\documentclass{pasj00}

\begin{document}
\SetRunningHead{R. Fujimoto et al.}{Warm-Hot Intergalactic Medium 
around the Virgo Cluster}
\Received{2004/05/09}
\Accepted{2004/08/10}

\title{Probing Warm-Hot Intergalactic Medium Associated with the Virgo
Cluster using an Oxygen Absorption Line}

\author{Ryuichi \textsc{Fujimoto}, Yoh \textsc{Takei}, 
Takayuki \textsc{Tamura},\\
Kazuhisa \textsc{Mitsuda}, and Noriko Y. \textsc{Yamasaki}
}
\affil{Institute of Space and Astronautical Science, 
Japan Aerospace Exploration Agency,\\
3-1-1 Yoshinodai, Sagamihara, Kanagawa 229-8510}
\email{fujimoto@astro.isas.jaxa.jp}

\author{Ryo \textsc{Shibata}}
\affil{Department of Physics, Nagoya University, Furo-cho, 
Chikusa-ku, Nagoya 464-8602}

\author{Takaya \textsc{Ohashi} and Naomi \textsc{Ota}\thanks{Present
Address: Cosmic Radiation Laboratory, RIKEN, 2-1 Hirosawa, Wako, 
Saitama 351-0198}}

\affil{Department of Physics, Tokyo Metropolitan University, 
1-1 Minami-Osawa, Hachioji, Tokyo 192-0397}

\author{Michael D. \textsc{Audley}}
\affil{UK Astronomy Technology Centre, Royal Observatory, 
Blackford Hill, Edinburgh, EH9 3HJ, UK}

\and

\author{Richard L. \textsc{Kelley} and Caroline A. \textsc{Kilbourne}}
\affil{NASA Goddard Space Flight Center, Greenbelt MD 20771, USA}




\maketitle

\begin{abstract}

To detect a warm-hot intergalactic medium associated with the
large-scale structure of the universe, we observed a quasar behind the
Virgo cluster with XMM-Newton.  With a net exposure time of 54~ks, we
marginally detected an O\emissiontype{VIII} K$\alpha$ absorption line
at $650.9^{+0.8}_{-1.9}$~eV in the RGS spectra, with a statistical
confidence of 96.4\%. The observed line center energy is consistent
with the redshift of M87, and hence the absorber is associated with
the Virgo cluster. From the curve of growth, the O\emissiontype{VIII}
column density was estimated to be $\gtrsim 7\times
10^{16}$~cm$^{-2}$. In the EPIC spectra, excess emission was found
after evaluating the hot intracluster medium in the Virgo cluster and
various background components. We inspected the ROSAT All-Sky Survey
map of the diffuse soft X-ray background, and confirmed that the level
of the north and west regions just outside of the Virgo cluster is
consistent with the background model that we used, while that of the
east side is significantly higher and the enhancement is comparable
with the excess emission found in the EPIC data. We consider a
significant portion of the excess emission to be associated with the
Virgo cluster, although a possible contribution from the North Polar
Spur cannot be excluded. Using the column density and the emission
measure and assuming an oxygen abundance of 0.1 and an ionization
fraction of 0.4, we estimate that the mean electron density and the
line-of-sight distance of the warm-hot gas are $\lesssim 6\times
10^{-5}$~cm$^{-3}$ and $\gtrsim 9$~Mpc, respectively. These numbers
strongly suggest that we have detected a warm-hot intergalactic medium
in a filament associated with the Virgo cluster.

\end{abstract}

\section{Introduction}

From several independent measurements --- cosmic microwave background
anisotropy analysis \citep{Spergel03}, primordial
deuterium-to-hydrogen ratio \citep{Burles01}, and Ly$\alpha$ forest
absorption at high redshift ($z \sim 3$) \citep{Rauch97} --- the
cosmic baryon density is now converging to $\Omega_{\rm b}\approx
0.04h^{-2}_{70}$, where $h_{70} = H_0/70$~km\,s$^{-1}$\,Mpc$^{-1}$ and
$H_0$ is the Hubble constant.  The local baryon density appears,
however, to be far lower than these indications (e.g.,
\cite{Fukugita98}; \cite{Bristow94}; \cite{Persic92}). Recent
large-scale galaxy formation simulations predict that most of the
baryons ($\sim 50$\%) at the present time reside in a warm-hot
intergalactic medium (WHIM) with temperatures of $10^5$--$10^7$~K, and
form filamentary structures associated with clusters and groups of
galaxies (e.g., \cite{Cen99}). This warm-hot gas is heated primarily
by shock heating of gas accreting onto the large-scale structure
\citep{Dave01}, and its fraction is increasing with time.  Therefore,
detecting those warm-hot baryons is very important, not only for
settling the `missing baryon' problem, but also for understanding the
formation of large-scale cosmological structure.

Considering the elemental abundance and the ionization fraction,
O\emissiontype{VII} and O\emissiontype{VIII} absorption lines are the
best tracers for probing the warm-hot gas at $10^6$--$10^7$~K
(\cite{Perna98}; \cite{Fang00}).  In fact, these lines have been
detected with grating spectrometers onboard Chandra and XMM-Newton,
toward PKS~2155--304, 3C~273, Mrk~421, and 3C~120 (\cite{Nicastro02};
\cite{Fang02}; \cite{Rasmussen03}; \cite{Fang03}; \cite{McKernan03}). 
In most cases, the center energy is consistent with a zero redshift,
which suggests detection of the WHIM in the local group.  However, if
its redshift is zero, distinguishing it from the warm-hot interstellar
matter in our Galaxy is not trivial \citep{Futamoto04}. On the other
hand, soft X-ray emission in clusters of galaxies has been studied
(e.g., \cite{Kaastra03}; \cite{Finoguenov03}). These authors claimed
the detection of a warm-hot component with $kT\sim 0.2$~keV.  However,
since the energy band below 0.5~keV is significantly contaminated by
emission from the Milky Way halo and the Local Hot Bubble, a precise
determination of the quantities of the warm-hot gas is very difficult
with the limited energy resolution of proportional counters and X-ray
CCDs.

In searching for a breakthrough, we decided to probe the warm-hot gas
{\it associated with nearby clusters of galaxies} through absorption
lines by observing a quasar behind the Virgo cluster with the
Reflection Grating Spectrometer (RGS; \cite{Herder01}) onboard
XMM-Newton \citep{Jansen01}. In this paper, we report on our
observation and results. We assume $h_{70}=1$ throughout this paper.
Errors are quoted at the 90\% confidence level, unless otherwise
stated.

\section{Sample Selection and Observation}

\begin{figure}
\begin{center}
\FigureFile(80mm,80mm){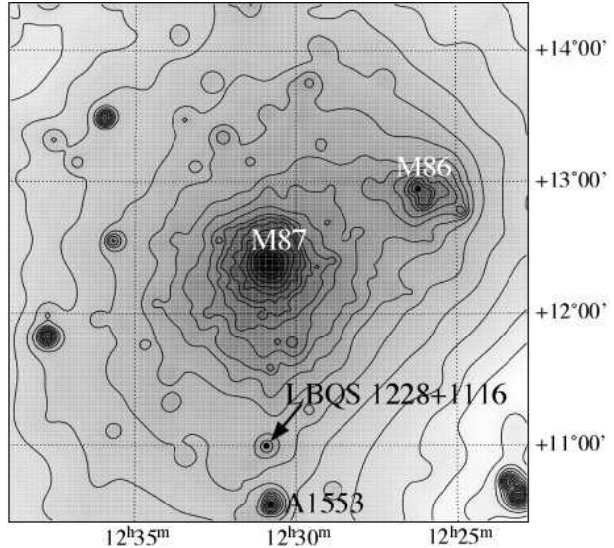}
\end{center}
\caption{ROSAT PSPC image around M87. The target, LBQS~1228+1116,
is located $83.3'$ south of M87.}
\label{fig:ROSAT_image}
\end{figure}

To find an appropriate target, we searched the ROSAT All-Sky Survey
(RASS) bright source catalogue \citep{Voges99} and the source
catalogue from pointed PSPC
observations\footnote{$<$http://www.xray.mpe.mpg.de/cgi-bin/rosat/rosat-survey/$>$}
for bright quasars behind the Virgo and Coma clusters.  The target
that we selected was LBQS 1228+1116. The location of this quasar is
(\timeform{12h30m54s.1}, \timeform{11D00'11''}), which is
\timeform{83.3'} (or 0.45~Mpc in linear scale) south of M87
(figure~\ref{fig:ROSAT_image}).  Because the redshift is $z = 0.237$
\citep{Hewett95}, it is located well behind the Virgo cluster. The
ROSAT PSPC count rate was 0.12--0.15~cps.

The observation was performed with XMM-Newton, from 2003 July 13 to
July 14. The observation ID is 0145800101. The instrument mode,
filter, and the net exposure time are summarized in
table~\ref{tab:obslog}. Note that, although 100~ks was scheduled, the
net observation time was only $\sim 50$~ks due to background flares.

\begin{table}
\caption{Observation mode and net exposure time.}
\label{tab:obslog}
\begin{center}
\begin{tabular}{llll}\hline\hline
Instrument & Mode       & Filter        & Net exposure\\\hline
RGS     & Spectrum+Q    & ---           & 53.8~ks\\
EPIC pn & Full frame    & Thin1         & 44.9~ks\\
EPIC MOS& Full frame    & Medium        & 56.9~ks\\\hline
\end{tabular}
\end{center}
\end{table}

\section{Data Analysis and Results}

\subsection{Absorption Line}

The observation data file (ODF) was divided into two files. We
performed data reduction using the XMM-Newton Science Analysis System
(SAS), version 5.4.1, with standard parameters. For the RGS, we checked
the data from a source-free region of CCD 9 for background flares,
and accumulated the signal photons only when the count rates in this
region were less than 0.15~cps for both RGS1 and 2. The net exposure
time was 8.8~ks for the first file and 45.0~ks for the second.  The
background spectra were produced using the same data sets. All of the
spectra were binned in four-channel bins (0.042~\AA), and the first and
second halves were summed.

\begin{figure}
\begin{center}
\FigureFile(80mm,80mm){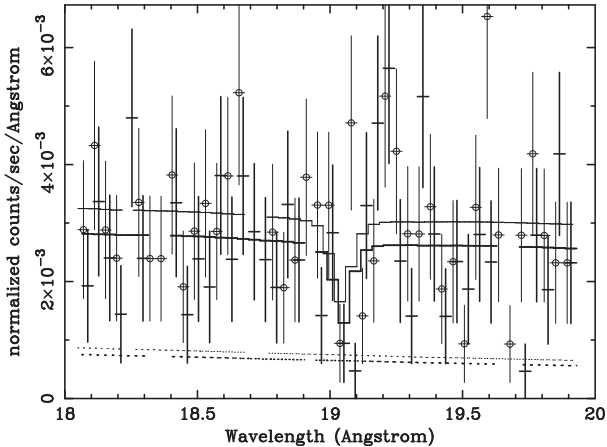}
\end{center}
\caption{Spectra around the O\emissiontype{VIII} K$\alpha$ line
obtained with RGS1 and 2. The cross and circle symbols represent the
data points of RGS1 and 2, respectively. The thick and thin solid
histograms show the best-fit models for RGS1 and 2. The backgrounds
were not subtracted, but were instead simultaneously fitted with a
constant model. The background levels are shown with dotted curves.
For illustrative purposes, channels with flickering pixels were
masked.}
\label{fig:RGS_spec}
\end{figure}

There was no sign of an O\emissiontype{VII} absorption line at around
21.6~\AA, where only the RGS1 operates. The 99.7\% upper limit of the
equivalent width was 2.8~eV at 21.69~\AA\ (571.6~eV).  Therefore, we
concentrated on the O\emissiontype{VIII} region.
Figure~\ref{fig:RGS_spec} shows the spectra obtained with the RGS1 and
2, between 18~\AA\ and 20~\AA. There is an absorption-line feature at
around 19.1~\AA\ in both RGS1 and 2.  We fitted the spectra with a
power law and a negative Gaussian, multiplied by the Galactic
absorption ($N_{\rm H} = 2.15\times 10^{20}$~cm$^{-2}$), using XSPEC
version 11.2. Since the number of photons, especially near the
absorption line, is small, we used the C-statistic (maximum-likelihood)
method.  When this method is used, a background spectrum cannot be
subtracted from the data. Therefore we estimated the background level
by fitting the background spectra of RGS1 and 2 simultaneously with a
constant model, and included it in the fitting model. The results are
summarized in table~\ref{tab:RGS_param}; the best-fit model is
shown in figure~\ref{fig:RGS_spec}. The line shape was consistent with
a narrow line, with a 90\% upper limit on a width of $\sigma
<5.1$~eV. Hence, the line width was fixed at 0.1~eV, which is small
compared with the detector resolution ($\Delta\lambda\sim
0.06$--0.07~\AA\ or $\Delta E\sim 2$~eV FWHM). When we calculated the
errors of the absorption-line parameters, we also fixed the photon
index of the power-law component. The observed line center energy and
equivalent width were $650.9^{+0.8}_{-1.9}$~eV and
$2.8^{+1.3}_{-2.0}$~eV, respectively. If we assume that this is an
O\emissiontype{VIII} absorption line, the energy shift from the rest
frame is $2.7^{+0.8}_{-1.9}$~eV, which corresponds to
$cz=1253^{+881}_{-369}$~km\,s$^{-1}$. This is consistent with that of
M87 ($cz=1307$~km\,s$^{-1}$).  Note that the systematic error in the
absolute energy scale of the RGS is 8~m\AA\ (rms) or 0.3~eV at 650~eV
\citep{Herder01}, which is much smaller than the statistical errors.

\begin{table}
\caption{Best-fit parameters of the absorption line.}
\label{tab:RGS_param}
\begin{center}
\begin{tabular}{ll}\hline\hline
\multicolumn{2}{c}{Absorption line}\\
\hline
Energy  & $650.9^{+0.8}_{-1.9}$~eV \\
Width ($\sigma$)\footnotemark[*] & $< 5.1$~eV (90\%)\\
Normalization 
& $(3.9^{+2.0}_{-2.8})\times 10^{-6}$~photons\,cm$^{-2}$\,s$^{-1}$ \\\hline
\multicolumn{2}{c}{Power law}\\\hline
Index           & 3.2 \\
Normalization\footnotemark[$\dagger$]
& $(3.57\pm 0.35) \times 10^{-4}$ \\\hline
C-statistic		& 177.02\\
d.o.f			& 167\\
\hline
\multicolumn{2}{@{}l@{}}{\hbox to 0pt{\parbox{85mm}{\footnotesize
\footnotemark[$*$] 
For all fits except for the determination of the statistical
error for this parameter, it was fixed at 0.1~eV.
\par\noindent
\footnotemark[$\dagger$] In units of photons keV$^{-1}$ cm$^{-2}$
s$^{-1}$ at 1~keV.
}\hss}}
\end{tabular}
\end{center}
\end{table}

We evaluated the statistical significance of the absorption line in the
following way. The best-fit model without an absorption line gives a
C-statistic of 181.76.  When we include an absorption line, but fix
the center and the width to the energy of the O\emissiontype{VIII}
K$\alpha$ resonant line with $cz=1307$~km\,s$^{-1}$ and
$\sigma=0.1$~eV, respectively, we obtain the best-fit C-statistic of
177.43.  Thus, the improvement, $\Delta C$, is 4.33.  The value of
$\Delta C$ approximately follows the $\chi^2$ statistic for 1 degree
of freedom if the absorption line is just a statistical fluctuation
\citep{Cash79}. In order to evaluate the statistical confidence
of the line detection, we produced 10000 simulation spectra using the
continuum model shown in table~\ref{tab:RGS_param} and the background
without the absorption line. We then fitted the spectra with and without
a Gaussian absorption or emission line model.  We obtained $\Delta C
\ge 4.33$ for 3.6\% of the simulation spectra. This is consistent
with the value expected from the $\chi^2$ distribution (3.7\%).
We thus conclude that the chance probability for detecting such an
absorption line at the very wavelength corresponding to
O\emissiontype{VIII} K$\alpha$ at the cluster redshift is 3.6\%; in
other words, the statistical confidence of the absorption line is
96.4\%.

The width of the wavelength range where both the RGS1 and 2 operate is
$\sim 15.6$~\AA.  Because the wavelength resolution is $\sim
0.06$~\AA, there are about 260 independent wavelength bins.  The
probability for detecting an emission/absorption structure of $\Delta
C \ge 4.33$ at {\it some} wavelength is almost unity with an expected
occurrence of 9.4. We searched for such structures in the observed
spectra, and actually found seven such structures other than that at
19.05~\AA.  However, none of them can be identified with any atomic
emission/absorption line with a reasonable oscillator strength at
$z=0$, the cluster redshift, or the quasar redshift.  Thus, these are
all considered to be statistical fluctuations, and only the 19.05~\AA\
feature can be real at a statistical confidence of 96.4\%.

We then investigated the systematic errors.  First, we checked the RGS
spectra of bright X-ray sources in the archival data to ensure that
there is no instrumental feature at that wavelength.  Then, since the
estimation of an absorption line could be influenced by the
determination of the continuum, we examined the dependence of the
equivalent width on the wavelength region used in the fits and the
choice of the continuum model.  For this purpose, we changed the width
of the wavelength band for fits from 1.0 to 2.8~\AA\ and added a
fifth-order polynomial to the continuum.  We found that, as long as
the width of the fitting region is wider than 1.6~\AA, the variation
of the equivalent width is smaller than 0.2~eV.  When the fitting
region is too narrow, the continuum level is not well constrained, and
is strongly coupled to the equivalent width of the line.

Assuming that the absorption line is not saturated, the equivalent
width is related to the ion column density in the following way:
\begin{equation}
EW 
= \frac{\pi h e^2}{m_{\rm e} c}f_{\rm os} \frac{N_{\rm ion}}{1+z},
\end{equation}
where $N_{\rm ion}$ is the column density of the ion and $f_{\rm os}$
is the oscillator strength of the transition \citep{Sarazin89}. We
used $f_{\rm os}=0.70$ for the O\emissiontype{VII} resonance absorption
line and $f_{\rm os}=0.42$ for that of O\emissiontype{VIII}
\citep{Verner96}.  Then, the O\emissiontype{VIII} column density was
$(6.2^{+3.3}_{-4.4})\times 10^{16}$~cm$^{-2}$, while the 99.7\%
upper limit on the O\emissiontype{VII} column density was $3.7\times
10^{16}$~cm$^{-2}$.

\subsection{Excess Emission}

We also analyzed the EPIC data, and searched for diffuse
emission. Since we are mostly interested in the low-energy band, the
pn detector is suitable for our analysis. Hence, we describe the
results obtained from the pn data here. We confirmed that the MOS data
gave similar results, but with larger errors.

\begin{figure}
\begin{center}
\FigureFile(80mm,80mm){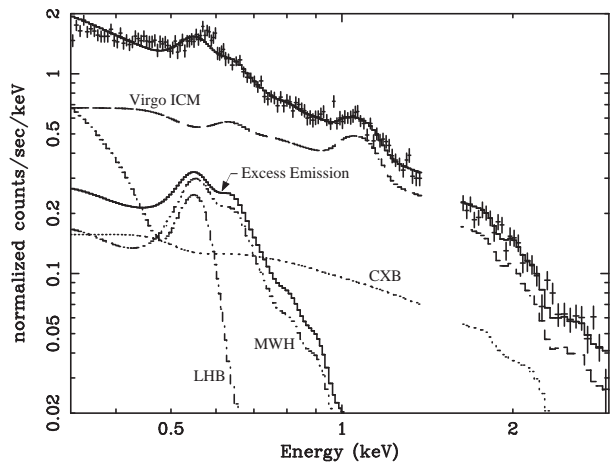}
\end{center}
\caption{Soft X-ray spectrum of the diffuse emission obtained with
EPIC pn. The aluminum line region was masked. Internal background was
subtracted using data taken with a filter wheel in the closed
position.  The parameters of the cosmic X-ray background (CXB), Milky
Way halo (MWH), and Local Hot Bubble (LHB) were fixed to those
obtained by \citet{Lumb02}. The data require another component that
can be modeled with an optically thin thermal plasma with
$kT=0.21$~keV.}
\label{fig:EPIC_spec}
\end{figure}

The number of events above 10~keV is regarded as a good indicator of
the internal background. Thus, we discarded the data where the rate of
pattern 0 events in this energy range was larger than 0.7~cps. Then, we
excluded LBQS~1218+1116 and other point sources using SAS
edetect\_chain, and accumulated photons from the diffuse emission in
the 0.3--3~keV band over the field of view.  To remove any
contribution from the internal background, we subtracted data taken
with the filter wheel in the closed position from the spectrum. 
We then modeled it with four components: hot intracluster medium (ICM) of
the Virgo cluster, cosmic X-ray background (CXB), Milky Way halo
(MWH), and Local Hot Bubble (LHB). For the CXB, MWH, and LHB, we used
parameters obtained by \citet{Lumb02} as a template. When we fixed the
temperature of the MWH (0.20~keV), the temperature of the LHB
(0.07~keV), and all of the CXB parameters, the temperature of the ICM
determined by the fit became $2.0^{+0.2}_{-0.1}$~keV, which is
consistent with that obtained with ASCA ($2.14\pm 0.12$~keV,
\cite{Shibata01}). The normalization of the MWH was, however, a factor
of 2.3-times larger than that obtained by \citet{Lumb02}. From an
analysis of eight data sets obtained with XMM-Newton, \citet{Lumb02}
indicated that the mean deviation of 0.2--1~keV flux is about 35\%
from field to field. Therefore, the level of the warm-hot emission
around LBQS~1228+1116 is significantly higher than that of the typical
Galactic background, even if we consider its fluctuation. We fixed the
parameters of the MWH, LHB, and CXB to those obtained by
\citet{Lumb02}, added another optically-thin thermal plasma component,
and fitted the entire spectrum. Note that the normalization of the
additional component ($N$) strongly couples with the oxygen abundance
($A$), and $A\times N$ is a good parameter. Here, we assumed
$A=0.1$. The results are summarized in table~\ref{tab:EPICpn}, and the
best-fit models are shown in figure~\ref{fig:EPIC_spec}.  Note that
the errors given in table~\ref{tab:EPICpn} do not contain the
uncertainties of the background components. The normalization of the
excess emission depends on the normalization, abundance, and
temperatures of the MWH and LHB, while the temperature of the excess
emission is sensitive to those of the MWH and LHB.

\begin{figure}
\begin{center}
\FigureFile(80mm,80mm){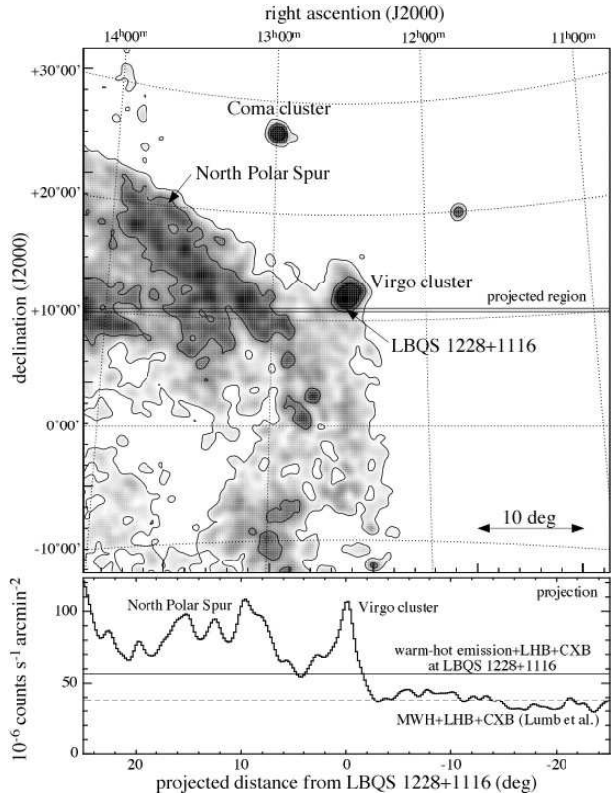}
\end{center}
\caption{(Upper panel) $\timeform{25D}\times\timeform{25D}$ image 
centered at LBQS~1228+1116, obtained from the RASS 3/4~keV map of the
diffuse X-ray background \citep{Snowden97}. The contours represent 80,
120, and $180\times 10^{-6}$~counts\,s$^{-1}$\,arcmin$^{-2}$. (Lower
panel) Projection of the $\timeform{0.5D}\times\timeform{50D}$ region
shown in the upper panel. The solid line represents the level of the
warm-hot emission plus LHB and CXB estimated from the present EPIC
data, while the dashed line represents the background level (MWH, LHB,
CXB) estimated from \citet{Lumb02}.}
\label{fig:RASS}
\end{figure}

It is known that the Virgo cluster is located close to the North Polar
Spur (Loop I) (e.g., \cite{Irwin96}). To evaluate its effect, we
inspected the RASS 3/4~keV map of the diffuse X-ray background
\citep{Snowden97}. The upper panel of figure~\ref{fig:RASS} shows a
$\timeform{25D}\times\timeform{25D}$ image centered at LBQS~1228+1116.
The lower panel is a projection of the
$\timeform{0.5D}\times\timeform{50D}$ region shown in the upper
panel. The levels of the background components (MWH, LHB, CXB)
estimated from the data obtained by \citet{Lumb02} and of the warm-hot
emission plus LHB and CXB estimated from the present EPIC pn data are
also shown in the panel with a dashed line and a solid line,
respectively. The level of the west side of the Virgo cluster
($\lesssim \timeform{-3D}$) is consistent with that of \citet{Lumb02},
while that of the east side ($\sim \timeform{+4D}$) is significantly
higher and comparable with the level of the warm-hot emission plus LHB
and CXB at LBQS~1228+1116.

\begin{table}
\caption{Best-fit parameters of the emission spectrum\footnotemark[$*$].}
\label{tab:EPICpn}
\begin{center}
\begin{tabular}{ll}\hline\hline
\multicolumn{2}{c}{Intracluster medium (ICM)}\\\hline
Temperature $kT$        & $2.17^{+0.14}_{-0.16}$~keV\\
Normalization $N$\footnotemark[$\dagger$]& $(3.0\pm 0.2)\times 10^{-3}$ \\
Abundance $A$           & $0.22^{+0.07}_{-0.05}$\\\hline
\multicolumn{2}{c}{Excess emission}\\\hline
Temperature $kT$        & $0.21\pm 0.01$~keV\\
Normalization $N$\footnotemark[$\dagger$]& $(1.5\pm 0.2)\times 10^{-3}$\\
Abundance $A$           &     0.1 (fixed)\\\hline
$\chi^2$	& 205.36\\
d.o.f.		& 146\\
\hline
\multicolumn{2}{@{}l@{}}{\hbox to 0pt{\parbox{85mm}{\footnotesize
\footnotemark[$*$] The parameters of the
background components were fixed at the values obtained by
\citet{Lumb02}. Errors do not contain systematic errors of the
background components.
\par\noindent
\footnotemark[$\dagger$] In units of $10^{-14}/4\pi (D_{\rm
A}(1+z))^2\int n_{\rm e}^2dV$, where $D_{\rm A}$ is the angular size
distance to the source (cm) and $n_{\rm e}$ is the electron density
(cm$^{-3}$).}\hss}}
\end{tabular}
\end{center}
\end{table}

\section{Discussion}
\label{sec:discussion}

Our results suggest the existence of a red-shifted
O\emissiontype{VIII} resonance absorption line with a statistical
confidence of 96.4\% due to limited statistics. The velocity shift is
$1253^{+881}_{-369}$~km\,s$^{-1}$, which is consistent with that of
M87 ($cz=1307$~km\,s$^{-1}$).  Thus, we claim detection of the WHIM
associated with the Virgo cluster. The equivalent width of the
O\emissiontype{VIII} K$\alpha$ line is $2.8^{+1.3}_{-2.0}$~eV. This is
much larger than those reported for WHIM in the local group
($EW=0.1$--0.4~eV; e.g., \cite{Rasmussen03}; \cite{Fang02}). The
intrinsic width of the line was not resolved, with an upper limit of
$\sigma<5.1$~eV (90\%). This corresponds to a Doppler $b$ parameter of
$<3300$~km\,s$^{-1}$. Because the Virgo cluster is thought to be
elongated along the line of sight by 12--30~Mpc \citep{Yasuda97}, a
velocity difference of 800--2100~km\,s$^{-1}$ is expected due to
cosmological expansion. On the other hand, a turbulent velocity of a
few hundred km\,s$^{-1}$ is expected for WHIM from simulations (e.g.,
\cite{Cen99}). From the curve of growth of the O\emissiontype{VIII}
K$\alpha$ absorption line (see \cite{Futamoto04}), the
O\emissiontype{VIII} column density $N_{\rm O\emissiontype{VIII}}$ is
estimated to be $6.8\times 10^{16}$, $9.6\times 10^{16}$, and
$2.4\times 10^{17}$~cm$^{-2}$, respectively, for $b=2100$, 800, and
400~km\,s$^{-1}$ with $EW=2.8$~eV.  $N_{\rm O\emissiontype{VIII}}\sim
1\times 10^{17}$~cm$^{-2}$ is close to the maximum column density
predicted by Fang and Canizares (2000), who investigated the column
density of the WHIM gas by Monte-Carlo simulations.

On the other hand, no O\emissiontype{VII} absorption line was
detected. From the upper limit, the ionization fraction ratio of
O\emissiontype{VIII} to O\emissiontype{VII} is constrained to be
$>1.7$, assuming that both of the O\emissiontype{VIII} K$\alpha$ and
O\emissiontype{VII} K$\alpha$ lines are not saturated.  If we further
assume collisional ionization equilibrium, this implies $kT> 0.20$~keV
\citep{Mazzotta98}. However, photoionization by the cosmic X-ray and
UV background radiation would increase the ionization fraction of
O\emissiontype{VIII} if the density is as low as $10^{-5}$~cm$^{-3}$
\citep{Chen03}. Since we can constrain only the upper limit of the
density (see next paragraph), this could be the case for the WHIM
around the Virgo cluster, and its temperature could be lower.

Excess emission was found in the EPIC spectra over the multi-component
model representing the emission from the Virgo ICM, CXB, LHB, and MWH.
The intensity of the multi-component model is consistent with the RASS
diffuse soft X-ray background map in the north and the west regions
just outside of the Virgo cluster. This strongly suggests that the
excess emission is associated with the Virgo cluster. However, a
possible contribution by emission from the North Polar Spur cannot be
excluded completely. We thus take the emission intensity as the upper
limit for the emission from the WHIM around the Virgo cluster. The
O\emissiontype{VIII} column density is related to the average electron
density, $n_{\rm e}$, and the line-of-sight length, $L$, of the WHIM
as $N_{\rm O\emissiontype{VIII}} = f A Z_{\odot} n_{\rm e}L$, where
$f$ is the ionization fraction, $A$ is the oxygen abundance relative
to the solar value, and $Z_{\odot}$ is the solar abundance of the
oxygen.  On the other hand, if we assume a uniform distribution of the
WHIM, the emission measure, ${\rm EM}$, of the plasma is ${\rm EM}
\sim n_{\rm e}^2 LS$, where $S$ is the geometrical area that we
observed, and is $7.3\times 10^{3}$~kpc$^{2}$ for the present data.
Then, the electron density and the line-of-sight distance are
constrained to
\begin{eqnarray}
n_{\rm e}
&\lesssim&6\times 10^{-5}~{\rm cm}^{-3}
\left(\frac{N}{1.5\times 10^{-3}}\right)
\left(\frac{A}{0.1}\right)
\left(\frac{f}{0.4}\right)\\ \nonumber
&&
\left(\frac{N_{\rm O\emissiontype{VIII}}}
{6.2\times 10^{16}~{\rm cm}^{-2}}\right)^{-1},\\
L
&\gtrsim &9~{\rm Mpc}
\left(\frac{N}{1.5\times 10^{-3}}\right)^{-1}
\left(\frac{A}{0.1}\right)^{-2}
\left(\frac{f}{0.4}\right)^{-2} \\ \nonumber
&&
\left(\frac{N_{\rm O\emissiontype{VIII}}}
{6.2\times 10^{16}~{\rm cm}^{-2}}\right)^{2},
\end{eqnarray}
where $N$ is the normalization of the excess emission (see
table~\ref{tab:EPICpn}), and 0.4 is the ionization fraction for
$kT=0.21$~keV. The derived density corresponds to a baryon overdensity
of $\delta\lesssim 250$.  The depth of $\gtrsim 9$~Mpc is much larger
than the linear dimensions of the Virgo cluster in the sky, but is
consistent with the elongation of 12--30~Mpc suggested by Yasuda,
Fukugita, and Okamura (1997), which is understood as being a filament
of the large-scale hierarchical structure along the line of sight.
Note that the line broadening due to the Hubble flow of the 9~Mpc
line-of-sight distance corresponds to about 1.4~eV. This is much
smaller than the 90\% upper limit of the line width obtained from the
present data. However, it could be resolved if the statistics were
good enough, since it is comparable to the energy resolution of the
RGS ($\sim 2$~eV FWHM).

In summary, the present results strongly suggest the presence of the
WHIM in a filament associated with the Virgo cluster at the 96.4\%
confidence level. To further constrain the physical parameters of the
WHIM, a precise determination of the absorption line profile is an
important next step.  Future high-resolution spectroscopic
observations of the emission are also strongly desirable for resolving
the Galactic and extragalactic components.

\bigskip
We acknowledge Dr. Yuichirou Ezoe, who helped in the analysis of the
RASS data. This work was supported in part by Grants-in-Aid for
Scientific Research by MEXT/JSPS (14204017, 12440067, and 15340088).








\end{document}